\documentclass [aps,prb,superscriptaddress,twocolumn,showpacs,a4paper,unsortedaddress,amsmath,amssymb,byrevtex,floatfix]{revtex4}

\usepackage[latin1]{inputenc}
\usepackage{graphicx}
\usepackage{dcolumn}
\usepackage{bm}
\usepackage{hyperref}
\usepackage{times}

\begin{document}

%%%%%%%%%%%%%%%%%%%%%%%%%%%%%%%%%%%%%%%%%%%%%%%%%%%%%%%%%%%%%%%%%%%%%
\author{V\'{\i}ctor M Garc\'{\i}a-Su\'arez}
\affiliation{Departamento de F\'{\i}sica, Universidad de Oviedo, 33007 Oviedo, Spain}
\affiliation{Nanomaterials and Nanotechnology Research Center CINN, CSIC-Universidad de Oviedo, Spain}

\author{Amador Garc\'{\i}a-Fuente}
\affiliation{Departamento de F\'{\i}sica, Universidad de Oviedo,  33007 Oviedo, Spain}

\author{Diego J. Carrascal}
\affiliation{Departamento de F\'{\i}sica, Universidad de Oviedo, 33007 Oviedo, Spain}
\affiliation{Nanomaterials and Nanotechnology Research Center CINN, CSIC-Universidad de Oviedo, El Entrego, Spain}

\author{Enrique Burzur\'{\i}}
\affiliation{Kavli Institute of Nanoscience, Delft University of Technology, PO Box 5046, 2600 GA Delft, The Netherlands}
\affiliation{IMDEA Nanoscience, Ciudad Universitaria de Cantoblanco, C/Faraday 9, 28049 Madrid, Spain}

\author{Max Koole}
\affiliation{Kavli Institute of Nanoscience, Delft University of Technology, PO Box 5046, 2600 GA Delft, The Netherlands}

\author{Herre S. J. van der Zant}
\affiliation{Kavli Institute of Nanoscience, Delft University of Technology, PO Box 5046, 2600 GA Delft, The Netherlands}

\author{Maria El Abbassi}
\affiliation{Department of Physics, University of Basel, Klingelbergstrasse 82, CH-4056, Basel, Switzerland}
\affiliation{Empa, Swiss Federal Laboratories for Materials Science and Technology, Uberlandstrasse 129, CH-8600 Dubendorf, Switzerland}

\author{Michel Calame}
\affiliation{Department of Physics, University of Basel, Klingelbergstrasse 82, CH-4056, Basel, Switzerland}
\affiliation{Empa, Swiss Federal Laboratories for Materials Science and Technology, Uberlandstrasse 129, CH-8600 Dubendorf, Switzerland}

\author{Jaime Ferrer}
\affiliation{Departamento de F\'{\i}sica, Universidad de Oviedo, 33007 Oviedo, Spain}
\affiliation{Nanomaterials and Nanotechnology Research Center CINN, CSIC-Universidad de Oviedo, El Entrego, Spain}

\email{ferrer@uniovi.es; enrique.burzuri@imdea.org}

%%%%%%%%%%%%%%%%%%%%%%%%%%%%%%%%%%%%%%%%%%%%%%%%%%%%%%%%%%%%%%%%%%%%%
\title{Spin signatures in the electrical response of graphene nanogaps}

%%%%%%%%%%%%%%%%%%%%%%%%%%%%%%%%%%%%%%%%%%%%%%%%%%%%%%%%%%%%%%%%%%%%%
\begin{abstract}
We analyse the electrical response of narrow graphene nanogaps in search for transport signatures originated from spin-polarized edge states. 
We find that the electrical transport across graphene nanogaps having perfectly defined zigzag edges does not carry any spin-related signature. 
We also analyse the magnetic and electrical properties of nanogaps whose electrodes have wedges that possibly occur in the currently fabricated 
nanogaps. These wedges can host spin polarized wedge low-energy states due to the bipartite nature of the graphene lattice. 
We find that these spin-polarized  low-energy modes give rise to low-voltage signatures in the differential conductance and to distinctive 
features in the stability diagrams. These are originated by fully  spin-polarized currents. 
\end{abstract}

\maketitle

\begin{figure*}
\includegraphics[width=\textwidth]{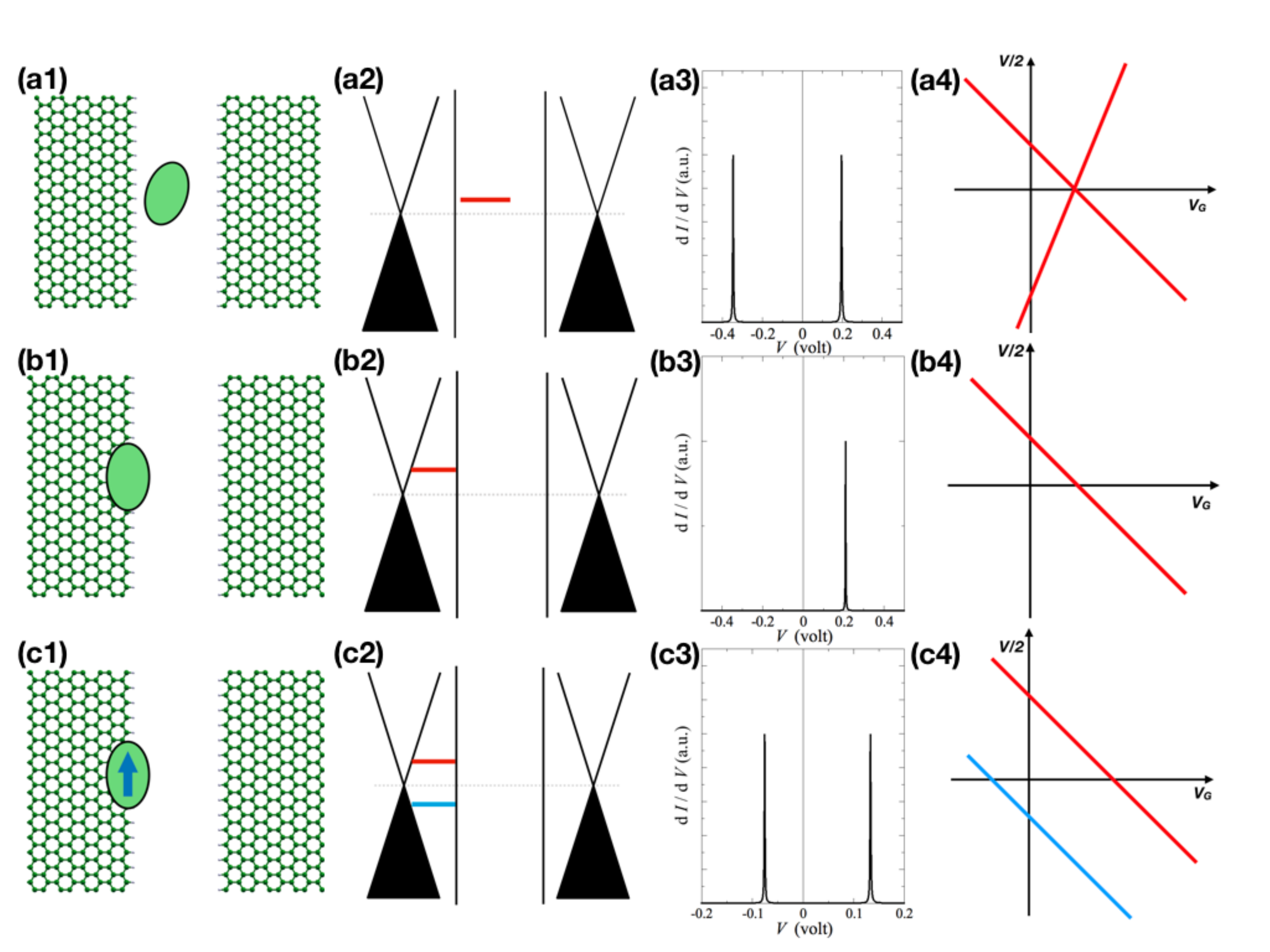}
\caption{Three possible realizations of a graphene nanogap junction displaying resonant tunnelling. We assume that the low-voltage current is
carried by only a single energy level as is usually the case for these junctions. (a1-a4) show a non-magnetic impurity located 
somewhere inside the  nanogap; (b1-b4) and (c1-c4) correspond to a non-magnetic or a magnetic defect placed at one of the two nanogap edges.
(a2), (b2) and (c2) are schematic  diagrams of the band structure and  energy levels at the nanogap. (a3), (b3) and (c3)
plot the corresponding differential conductance plots. Conductance peaks usually appear in the range 0.1-0.5 volt. 
(a4), (b4) and (c4) are the stability diagrams of the three junctions. We find in this
article that graphene wedges can be an important realization of the third scenario, where the spin-polarized level corresponds to a zero-energy
mode whose spin-degeneracy is split by the Coulomb Hubbard interaction. The splitting is usually not too large, that leads to conductance peaks
at about 50 to 200 mvolt.
 }
\label{fig:summary0}
\end{figure*}
\section{Introduction}
The development of graphene-based nano-electronics requires the fabrication of well-controlled nanometer-scale gaps. Graphene nanogaps 
with gap lengths of the order of one to few nanometres have indeed been realized in the past few years, via electro-burning 
techniques\cite{Pri11,Nef14,Geh15,Maria17}. Very recently, graphene nanogaps have been fabricated via the Mechanically 
Controlled Break Junction technique \cite{Caneva18}.  However, while these nano-gaps have indeed a length in 
the nanometer range, control over their morphology and content of the resulting edges is still in its infancy. Indeed, the
differential conductance of the nano-gaps presents frequently a low-voltage resonance peak that is also manifested as a cross in the 
stability  diagram (i.e.: a plot of the bias $V$ versus the gate voltage $V_G$), as we depict schematically in the first row of Figure \ref{fig:summary0}.
This resonant tunnelling behaviour is possibly due to molecular species left inside the nanogap by the electro-burning process. 
A possible question that may arise is whether localized electronic states at the edges can also lead to non-trivial tunnelling features. 
These features could only be measured above the electrical noise if the nanogap is as short as half a nanometer, or whether the localized
state protrudes enough from the edge. This scenario is depicted in the second and third rows of Figure \ref{fig:summary0}, that 
correspond to a non-magnetic and a magnetic state, respectively.  We analyse in this article situations where this third scenario may be realized.

\begin{figure}
\includegraphics[width=\columnwidth]{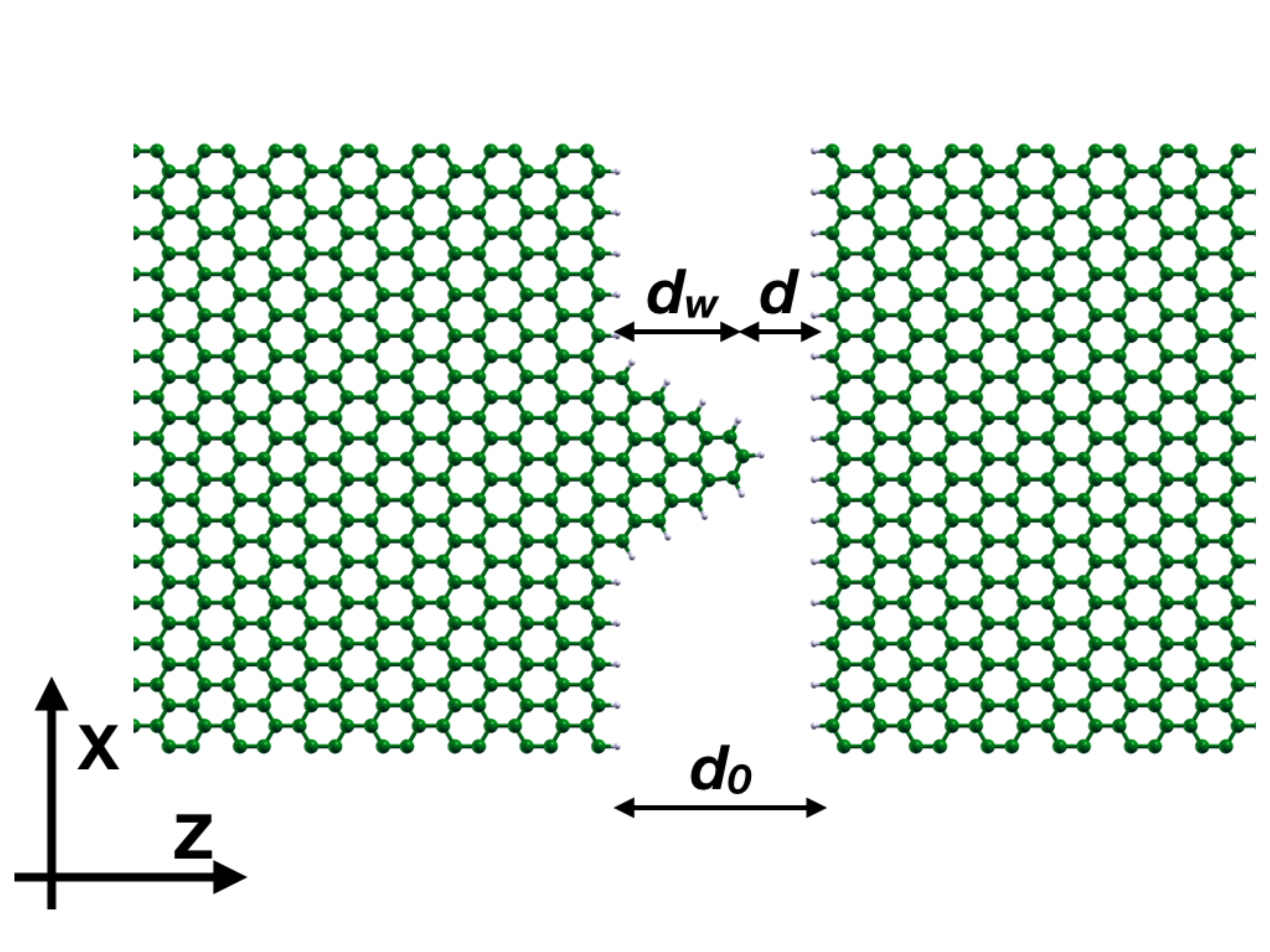}
\caption{\label{fig:wedge-example}
A graphene nanogap whose left sheet displays a wedge with lattice unbalance $I=4$. The figure shows the convention for coordinate axes used
in this article as well as the apparent nano-gap length $d_0$, the wedge height $d_w$ and the actual tunnelling length $d$.  The wedge
has a height $d_w \sim 8$ \AA. We assume that the wedges in this article are placed at the left sheet unless otherwise stated.}
\end{figure}

In the first situation, the graphene sheets at the nano-gap are terminated by hydrogenated zigzag graphene edges, that are predicted to 
carry spin polarized states with a local moment at the carbon edge atom aproaching 1/3 $\mu_\text{B}$
\cite{Fujita96,Leh03,Son06,PCMH07,Yaz07,Yaz10,Li13,Meunier17}. Experimental evidence of the existence of these
spin-polarized  edge states has emerged only recently \cite{Louie11,Baringhaus14,Slota18} 
In the second and more realistic situation, we assume that one of the edges shows a protuberance in the form of a wedge, as 
we depict in Figure \ref{fig:wedge-example}. Vacancies, voids,  islands of different shapes or wedge-terminated edges  
have also been predicted to display local magnetic moments\cite{Lie89,Ros07,PFB08,Car12,Lad16}. Spin-polarized  localized states in graphene 
sheets have been recently demonstrated via  hydrogen decoration\cite{Gon16}, while nano-scale graphene islands passivated with hydroxyl 
groups  have also been found to be  magnetic \cite{SZPCZFZTD17}.  We analyse in this article simple hydrogen passivation as a representative 
example, although we will also show some results for hydroxyl passivation of isolated graphene islands.
 
Magnetism at graphene edges can arise if the dangling $\sigma$ bonds of the edge carbons are saturated with chemical 
groups so that they retain the bipartite nature of the original lattice and its half-filling nature\cite{PY12}. In addition, the p$_z$ orbitals at the
edges must remain not fully saturated. This can be achieved not only by simple hydrogen passivation, but also by passivation
via hydroxyl, carboxyl and COO groups\cite{SZPCZFZTD17,ZYXXG14}. Under these conditions, the sheets can be described to a 
first approximation by the half-filled Hubbard model on a bipartite lattice and Lieb's theorem applies\cite{Lie89}. This theorem states 
that if a bipartite lattice has $N_\text{A}$ and $N_\text{B}$ A and B sites so that there is a net unbalance $I=N_\text{A}-N_\text{B}$, then 
the ground state has a moment $m=2\,S_z=I$. Additionally,  the energy spectrum displays $I$  zero-energy modes, that are spin-split 
symmetrically about the Fermi energy by an amount $\Delta$ that is proportional to the on-site Coulomb repulsion 
$U$ \cite{PFB08,M15,laszlobook}. The occupied spin-up modes give rise to the finite moment $m$ predicted by Lieb's theorem.

We analyse first nanogaps containing two hydrogenated zigzag edges, that carry spin polarized edge states, that we call in this
article ZZ nanogaps. We find that the magnetic anisotropy of the spin-polarized edge states is consistent with a zero value and that 
their exchange 
coupling across the gap is minute and has a very short decay length. In consequence, the edge spin states show Heisenberg 
super-paramagnetic behaviour at any relevant laboratory temperature. This means that graphene nanogaps cannot show 
magnetorresistive behaviour. We also find that the differential conductance across these ZZ nanogaps is featureless and therefore does
not carry any spin signature. We also find that the tunnelling decay length across the gap $d_t \sim 0.4$ \AA, i.e.: is smaller than the 
Bohr radius. This important fact indicates that the electrical transport across a short-length nanogap must be dominated by those protrusions
having a height $d_w$ larger than about $10 \,d_t \sim 4$ \AA. 

This is the reason why we turn to analyse in the second place nanogaps where one electrode displays a protrusion or 
wedge with a sub-lattice unbalance $I$ equal or larger than three or four, as is the case of Figure \ref{fig:wedge-example}. 
These are called WZ nanogaps in the present article. We find, in  agreement with previous work \cite{Ros07,PFB08,Car12}, that these 
electrodes display a magnetic moment localized at the  wedge whose magnitude obeys Lieb's theorem\cite{Lie89}. We find low 
energy, single-electron spin-polarized states localized at the wedge. One of these spin-polarized states is placed at the wedge tip and 
gives rise to a single pair of low-voltage peaks in the differential conductance, as shown in bottom row of Figure \ref{fig:summary0}. 
We  finally analyse WW nanogaps where two protrusions are roughly facing each other. These nanogap configurations display a more complex
electrical response. We however expect that most experimental nanogaps should be of the WZ kind because the very short tunnelling
decay length. 

We show finally some experimental differential conductance d$I$/d$V$ traces that we have measured across several  nanometer-size gaps,
that show a pair of peaks.  These measurements are consistent with the response of the WZ nanogaps discussed in the article, but also
with the more conventional resonant tunnelling scenario shown in the top row of Figure \ref{fig:summary0}. 
We expect that the technology for nanogap fabrication will undergo a substantial development in the near future, and that 
control over the morphology of graphene edges as well as edge decoration may be attained in due time. We therefore propose the mechanism
laid in this article as a simple means to produce fully spin polarized electrical currents in a nanogap setting.  

\section{Results and Discussion}

\subsection{Electrical response of a ZZ nanogap}
\begin{figure}
\includegraphics[width=\columnwidth]{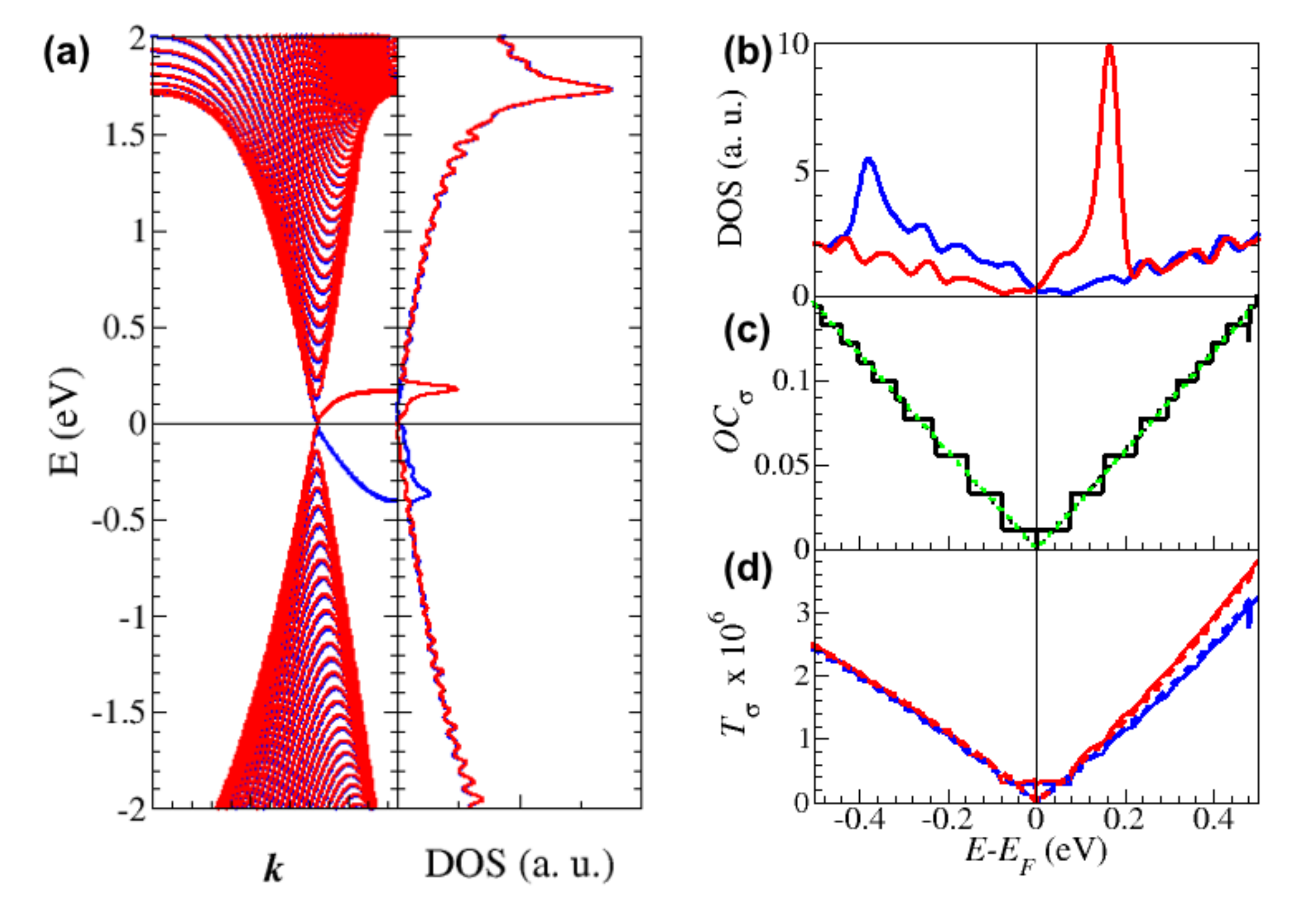}
\caption{\label{fig:ribbons_DOS}
(a) Spin-polarized bulk and edge bands, and spin-polarized Density of States  (DOS) of a 
zigzag ribbon having a width of $N_z=40$ unit cells. Blue and red lines indicate spin-up and down components, 
respectively. We set all DOS axes in arbitrary units (a.u.) in this article.
(b) Zoom of the spin-polarized DOS to display more clearly the
edge states. The ripples are due to the finite number of transverse k-points used in the calculation ($n_k=180$). 
(c) Number of Open Channels as a function of the energy $OC(E)$.  Two calculations are shown: a non-converged calculation
showing steps corresponds to $n_k=180$ (solid black line);  a converged calculation having $n_k=1440$ does not show steps 
(dotted green line). (d) Spin-polarized transmission of zigzag nanogaps. The solid lines correspond to $n_k=180$ and show ripples. 
The dashed lines do not show ripples, and correspond to $n_k=1440$. 
}
\end{figure}

The electronic and magnetic properties of hydrogenated zigzag nanoribbons have been discussed extensively in the past. We have 
therefore relegated our Density Functional Theory calculations on the matter to the Supporting Information, together with a few new results, 
that we summarize here.
We find that the magnetic anisotropy energy $D$ of the edge zigzag states is of order or smaller than 0.01-1 $\mu$eV, actually consistent 
with a value of zero. 
We therefore expect that edge-state spin coherent dynamics be driven by coherent rotation processes. 
We show in Figure \ref{fig:ribbons_DOS} (a) the band structure and the Density of States of a zigzag ribbon that features two peaks corresponding 
to the  spin-polarized edge states. 

We make now a ZZ nanogap based on the above ribbon to compute the transport properties across the gap. 
We have first estimated the 
exchange interaction $J$ across the nanogap and found it to be as small as  
\begin{equation}
J_\text{gap}=J_0\,e^{-d/d_0}\approx 22\,e^{-d (\text{\AA})/3,8}\,\mathrm{\mu eV}
\end{equation} 
so that we expect that the spin alignment of the edge states at both sides of the junction will be fluctuating between parallel (P) and 
anti-parallel (AP) configurations at any laboratory temperature.
The DOS of each sheet is plotted in Figure \ref{fig:ribbons_DOS} (b) in a narrow energy window where we focus on the
peaks originated by the edge states, that are still there. 
The Open Channels $OC(E)$ are an intrinsic property of each electrode, and measure the available number 
of conduction channels at a given energy $E$ that the electrons can use to impinge onto a scattering region, i.e.: the gap in our case. Figure 
\ref{fig:ribbons_DOS} (c) shows that the $OC(E)$ for our graphene electrodes possess electron-hole symmetry and are spin-unpolarized. 
Figure \ref{fig:ribbons_DOS}(d) shows that the peaks in the DOS do not translate into peaks in the spin-resolved transmission 
$T_\sigma(E)$. This curve therefore 
demonstrates that the edge states do not contribute to the electrical transport across the gap. 
Furthermore, we have computed 
the spin-resolved transmission function $T_\sigma(E)$ for the parallel (P) and anti-parallel (AP) spin alignments of the edge states. We have found that 
$T_\sigma(E)$ is featureless for both spin alignments  in the energy window shown in the Figure. 
We have also found that $T_\sigma(E)$ for the AP alignment is
exactly the average of $T_\uparrow(E)$ and $T_\downarrow(E)$ for the P alignment, yielding a theoretical zero magnetorresistance ratio. 
The slopes of $T_\sigma(E)$ are different for positive and negative energies, reflecting a slight loss of electron-hole symmetry. 
This is remarkable because 
an inspection to the bulk graphene conduction channels reveals that these are electron-hole symmetric. 

\begin{figure*}
\includegraphics[width=\textwidth]{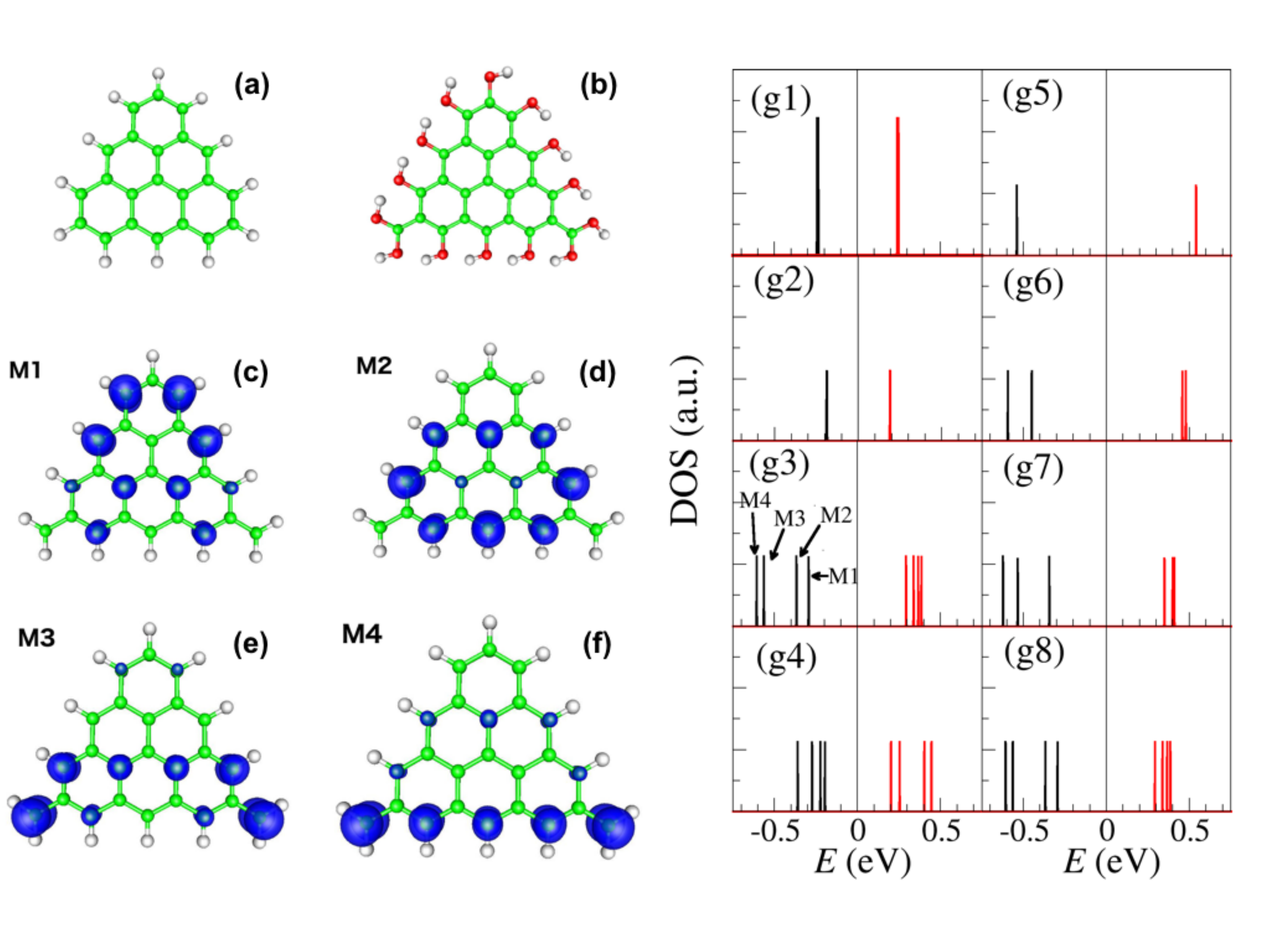}
\caption{\label{fig:islands} (a) Triangular island having $R=4$ rows and lattice unbalance $I=2$, 
passivated with hydrogen atoms (R4I2O). 
(b) Triangular island having $R=4$ rows and unbalance $I=4$, that is passivated by hydroxyl groups (R4I4OH).  
(c), (d), (e) and (f) Spin-up LDOS of the four spin-up low-energy modes of the  R4I4H island, M1-M4.
Spin-resolved DOS of (g1) R4I2H, (g2) R4I2OH, (g3)  R4I4H, (g4) R4I4OH, (g5) R1I1H, (g6) R2I2H, (g7) R3I3H and (g8) R4I4H islands. 
Solid black and red lines denote spin-up and -down  components. The peaks in (g1) are doubly-degenerate and therefore
are twice as high as the rest. Close inspection to (g2) reveals two almost degenerate peaks. }
\end{figure*}

The featureless differential conductance of a ZZ nanogap enables us to 
determine unambiguously the dependence of $G$  on the  gap length $d$, measured as the distance between the hydrogen atoms on 
both edges. We find that the zero-bias conductance shows the expected exponential decay, i.e. $G(d)=e^{- d/d_t}$, 
with a tunnelling decay length $d_t \approx 0.34$ \AA\, i.e.: smaller than a Bohr radius. 
This tiny decay length means that a ZZ nanogap should display negligible tunnelling current for $d\sim 1.5$ nm or larger. 
It also means that the nanogap electrical response should be dominated by the largest protrusion or wedge appearing at any of 
the two nanogap edges.
To place this statement on a firmer ground, we estimate the current flowing through a WZ nanogap where one of 
the two edges has a wedge having two extra rows.  This means
that the carbon atom at the wedge tip is $\sim 4$ \AA\ closer to the opposite electrode than the outermost carbon atoms at the adjacent 
zigzag segments. Using the above decay length, we estimate that the contribution of this tip atom to the tunneling current is 
$e^{4/d_\text{eff}} \sim 10^5$  times larger than the contribution of carbon atoms at the zigzag segments, meaning that the wedge
contribution dominates the gap conductance.  

\begin{figure*}
\includegraphics[width=\textwidth]{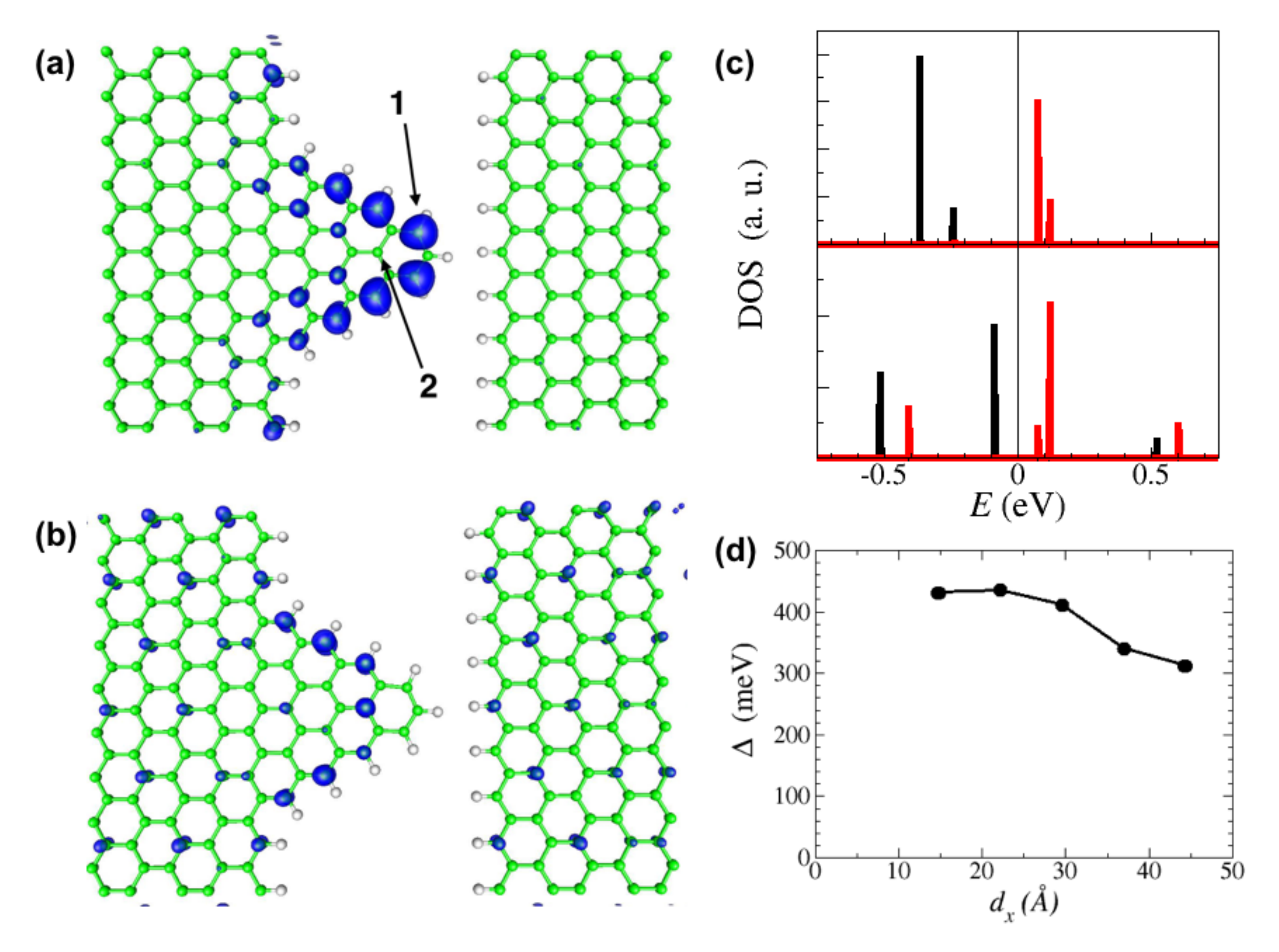}
\caption{\label{fig:rho_ws}WZ nanogap having a $R=4$ wedge, so that the wedge unbalance is $I=4$. 
The gap lengths are $d= 1$ nm and $d_\text{eff}=0.2$ nm. 
(a) and (b) LDOS of the spin-up M1 and M2  low-energy modes in a WS nanogap. (c) Spin-resolved DOS projected
into atoms 1 and 2. The peaks correspond to the modes M1 and M2 in Figure \ref{fig:islands}.
(d) Spin splittings $\Delta_\text{M1}(d_x)-\Delta_\text{M1}(\infty)$ for a wedge with $R=4$. }
\end{figure*}

\subsection{Properties of triangular islands}
We discuss now the properties of graphene islands having a shape similar to the wedges. We show in
Figure \ref{fig:islands} (a), (b) two islands that have  $R=4$ rows. They have an unbalance $I=2$ and $I=4$, respectively. 
We have passivated the two islands with hydrogen or with hydroxyl groups, and use the notation R4I2H,  
R4I2OH, R4I4H or R4I4OH, respectively. 
Similar structures with unsatured carbon p$_z$ orbitals at the edges were addressed a few years ago in 
references \onlinecite{Ros07,PFB08}, and we recapitulate here some rules relevant for the present work: (1) The structures
are well represented by the half-filled Hubbard model in a bipartite lattice and therefore satisfy Lieb's 
theorem\cite{Lie89}; (2) the single-particle 
spectrum of the structure has $I$ zero-energy states 
per spin channel, that are exchange split as $\epsilon_{i,\sigma}=\sigma\,\Delta_{i}/2,\,(i=1,...I$) ; 
(3) because there are $I$ electrons available, the resulting state has a moment $m=I$.
We have verified that the moment is $m=I$ for the four islands in accordance with Lieb's theorem \cite{Lie89}, and that each
island has $I$ modes having spin 1/2, each of them split by $\Delta_i$, as we show  in Figure \ref{fig:islands} (g1)-(g8), where we plot
the Density of States (DOS) of the islands in the neighbourhood of the Fermi level. We have placed the modes split
symmetrically about the Fermi energy to aid the eye and to connect with  the conclusions drawn for the Hubbard model \cite{Ros07}. 
The splitting depends on the kind of passivation: we find that hydroxyl groups yield a smaller
splitting than hydrogen atoms. In addition, this spin splitting is inversely proportional to the island's size, as we show in
the Figure (see also Ref. \onlinecite{PFB08}). We focus now in the R4I4H island (Figure \ref{fig:islands} (g3)), whose shape is closest to the $R=4$ wedges in this article. 
This island has four low-energy modes, that we call M1, M2, M3 and M4. We find that their degeneracy is broken because of island's lower symmetry,
as shown in Figure \ref{fig:islands} (g3). The four states of the 
spin-down DOS are quasi-degenerate, and the spin-splitting $\Delta_i$ is of $\sim 0.6 - 0.9$ eV. Figure 
\ref{fig:islands} (c)-(f) shows the spin-up Local Density of States (LDOS) of each mode M1-M4. They have a distinct shape and location across the 
island that  will be reflected later on in their specific contribution to the electrical transport across WS  gaps.
 \begin{figure*}
\includegraphics[width=\textwidth]{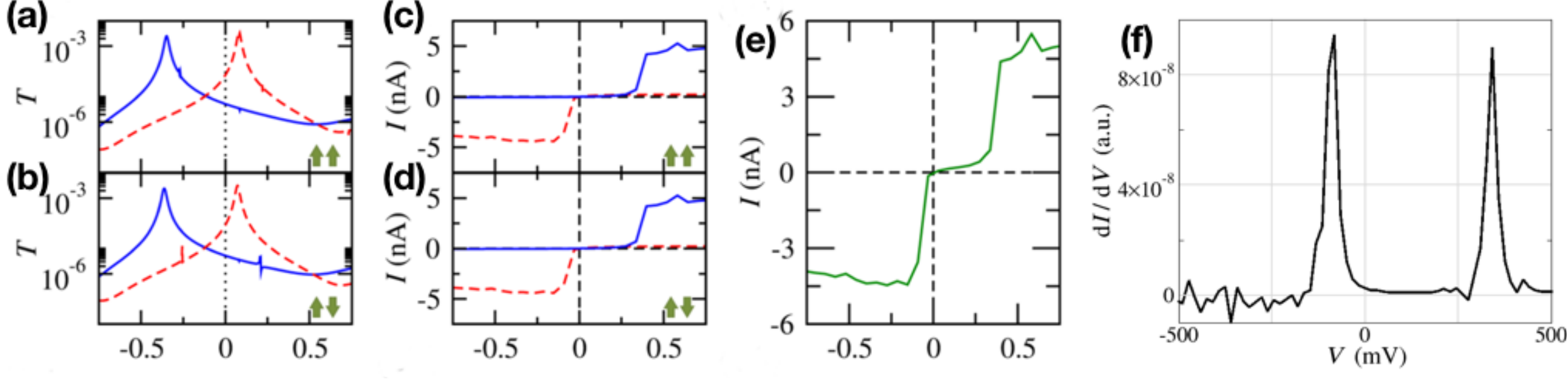}
\caption{\label{fig:WZ} Electrical properties of a WZ nanogap having a $R=4$ wedge, so that the wedge unbalance is $I=4$. 
The gap lengths are $d_0= 1$ nm and $d=0.2$ nm. 
(a) and (b) are the spin-resolved transmission $T_\sigma(E)$ for P and AP edge spin alignments, respectively; (c) and (d) are the 
spin-polarized currents as a function of voltage for the same orientations. (e) is the total electrical current as a function of voltage.
(f) is the differential conductance as a function of voltage.}
\end{figure*}

\subsection{Magnetism and zero-energy modes of graphene wedges}
We aim at describing now the properties of semi-infinite sheets whose single edge has a triangular wedge attached to it having $R$  rows
as in Figure \ref{fig:rho_ws} (a). This wedges are similar in shape to the islands in Figure \ref{fig:islands} (b). 
We set a nanogap length $d_0\sim 1$nm,  long enough to avoid any interaction across the nanogap.
By construction, the sublattice unbalance of these single-wedge nanogaps is $I=R$, and we 
have checked that our DFT calculations yield a ground state with $m=I$ as expected.
We analyse in detail the $R=4$ wedge for the sake of concreteness. The wedge hosts low-energy modes whose LDOS
look similar to those shown in Figure \ref{fig:islands} (c)-(f), as expected.
We plot the spin-resolved DOS projected onto atoms 1 and  2 at the wedge tip in Figure \ref{fig:rho_ws} (c) . We find that the low-energy 
PDOS at atom 1  is dominated by a single spin-split mode that we identify as the M1 mode of the equivalent $R=4$ island. To confirm this 
assertion, we show in Figure \ref{fig:rho_ws} (a) the spin-up LDOS
of the M1 mode, that looks extremely similar to that in Figure \ref{fig:islands} (c). It also shares a similar, somewhat smaller, 
spin-splitting $\Delta_\text{M1} \sim 0.45$ eV. The PDOS projected onto atom 2, that we show in Figure \ref{fig:rho_ws} (c), shows that the  
M2 mode splits into two peaks. This is due to hybridization with the bulk graphene modes.
This can be  seen by looking at the LDOS of each of the two split spin-up peaks. We show the one corresponding to the peak at -0.5 eV in 
Figure \ref{fig:rho_ws} (b). Because of this hybridization the spin-splitting is smaller, of order 0.1 eV. 
We find that the other two modes are localized away from the wedge tip, and closer to the semi-infinite sheet. 

However, the energy of the wedge modes is not placed symmetrically about the Fermi energy, losing the spin-resolved electron-hole symmetry.
For example, we find that $\epsilon_{M1,\uparrow}\simeq -0.35$ eV and $\epsilon_{M1,\downarrow}\simeq +0.10$ eV. 
This symmetry loss is physical and due to the electron-hole symmetry loss inherent in the DFT hamiltonian. We note that electron-hole
symmetry is lost in a bi-partite lattice by adding hopping or Coulomb interaction matrix elements linking atoms in the same sub-lattice
(i.e.: second nearest neighbours, see the extended Hubbard model discussed in the Supplementary Information).

We note that  a real electrode has many protrusions at its edge. 
As demonstrated above, we expect that one of these protrusions will dominate the tunneling current. For discussion, we assume
that the dominating wedge is a $R=4$ wedge. We find below that only the M1 mode gives rise to a low-voltage conductance peak in 
our transport  simulations, that appears  at a voltage $V=\Delta_\text{M1}/e$, so we focus from now on on this mode and on its 
spin-splitting.   We have found above that $\Delta_\text{M1}$ 
depends on the kind of carbon passivation, being larger for hydrogen than for hydroxyl passivation. We have also found that it is inversely 
proportional to the wedge size. However, $\Delta_\text{M1}$ may depend also  on the distance $d_x$ between the dominating wedge and its 
closest cousin.  We have estimated $\Delta_\text{M1}(d_x)$,  via our simulations of a WZ nanogap, where we vary the height 
$d_x=2.4\,N_x$ \AA\ of the unit cell. The bottom panel in Figure \ref{fig:rho_ws} shows our results for the discussed
$R=4$ wedge. This wedge has a length of about 12 \AA\ at the base, so the first few distances $d_x$ correspond to the "near field" 
interaction. While we do not have enough data points to attempt fitting the curve, we notice that the behaviour is consistent with a "far field" 
interaction law 
\begin{equation}
\Delta_\text{M1}(d_x)=\Delta_\text{M1}(\infty)+\frac{A}{d_x^n}
\end{equation}
where $\Delta_\text{M1}(\infty)\simeq 230$ meV and an exponent $n \sim 1-2$. 

\subsection{Electrical response of a WZ nanogap}
We discuss now the electrical response of WZ nanogaps, where we illustrate our results with the $R=4$ WZ nanogap shown in Figure 
\ref{fig:rho_ws}. We have set the gap length  $d=0.2$ nm so that the  apparent length is $d_0=1.0$ nm. We fix the  wedge spin in the up-direction, 
and orient the spin of the right-electrode zigzag edge upwards or downwards. Figures \ref{fig:WZ}  (a) and (b) show that the 
spin-resolved  transmissions for P and AP edge state  spin alignments of the two electrodes are exactly the same. 
Therefore, we demonstrate again that the spin-polarized edge state at the right zigzag edge does not contribute to the electrical transport. 
The Figure also shows a single low-energy spin-split resonance, that appears exactly at the energy $\epsilon_{M1,\sigma}$  for the
corresponding spin component. This corresponds to electron tunneling from the M1 mode at the wedge to the continuum of bulk states at
the right electrode. The other wedge modes are localized further inside the wedge and deliver a negligible contribution to the 
electrical current. We have simulated WZ nanogaps with different $R$, and have found that the height and width of the transmission 
resonance is proportional to the size of the wedge. 

\begin{figure*}
\includegraphics[width=\textwidth]{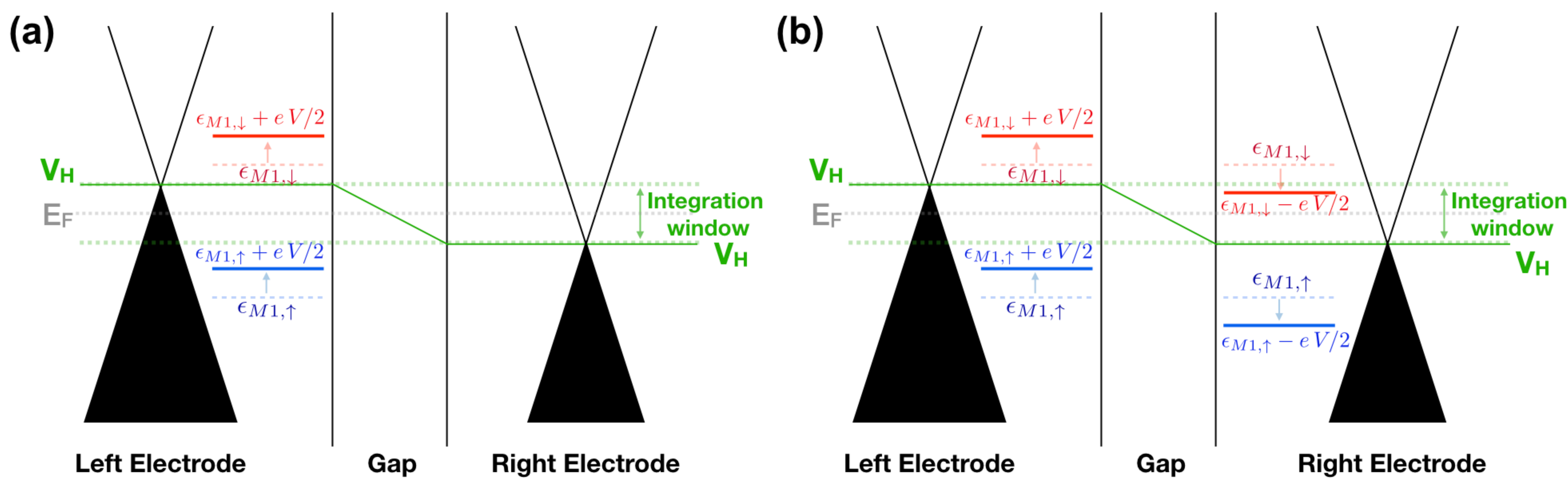}
\caption{\label{fig:diagram_ws}Energy level diagram for (a) a WZ nanogap and (b) a WW nanogap in a P configuration. 
The Dirac cones of each semi-infinite electrode are displayed at the extreme left and right of the figures. 
The Hartree potential $V_H$, and therefore the electrochemical potential, are shifted by 
the electrostatic potential originated by the bias voltage by the amount $\pm e\,V/2$. A similar shift is seen in the WS nanogap
for the M1 wedge mode at the left electrode $\epsilon_{M1,\sigma}\longrightarrow\epsilon_{M1,\sigma}+e\,V/2$.
In the WW nanogap, the M1 wedge modes at the left $L$ and right $R$ electrode shift in opposite directions: 
$\epsilon_{L,M1,\sigma}\longrightarrow\epsilon_{L,M1,\sigma}+e\,V/2$; $\epsilon_{R,M1,\sigma}\longrightarrow\epsilon_{R,M1,\sigma}-e\,V/2$.
The electrical current is calculated in terms of an integral of $T_\sigma(E)$ where the integration window at zero temperature extends to 
$\pm e\,V/2$. }
\end{figure*}

To understand the current-voltage curves, we draw the energy-level diagram shown in Figure \ref{fig:diagram_ws} (a). 
The spin-split resonances corresponding to the M1 mode are localized in the wedge, i.e.: at the left electrode. They are centred at 
$\epsilon_{M1,\sigma}$ at zero voltage and do not possess electron-hole symmetry as discussed above. These  two resonances shift up in 
energy when a positive bias voltage is applied according to  $\epsilon_{M1,\sigma} + eV/2$.
Therefore only the (negative-energy)  spin-up resonance enters the integration window $(-e\,V/2, e\,V/2)$ for positive bias. 
This happens at the threshold voltage  $V=|\epsilon_{M1,\uparrow}|/e$, resulting in a sudden jump on the current, that is 
fully spin-up polarized. The resonance stays inside the integration window upon further increase of the bias voltage, and
as a consequence the current remains constant. Conversely, only the (positive-energy) spin-down resonance enters the 
integration window at negative bias at $V=\epsilon_{M1,\downarrow}/e$ resulting in a   fully polarized spin-down current.
The resulting spin-resolved and spin-summed current-voltage characteristics are shown  in Figure 
\ref{fig:WZ} (c), (d) and (e), while the differential conductance is plotted in Figure \ref{fig:WZ} (f).  

We mention that the application of a magnetic field modifies the energy of these spin 1/2 modes by the usual amount 
$\pm g\,\mu_\text{B}\,B/2\simeq 0.058 \,B$ meV, where $B$ is measured in Tesla.
This means that a magnetic field as large as 10 T shifts the modes by an amount of 0.5 meV.

\begin{figure*}
\includegraphics[width=\textwidth]{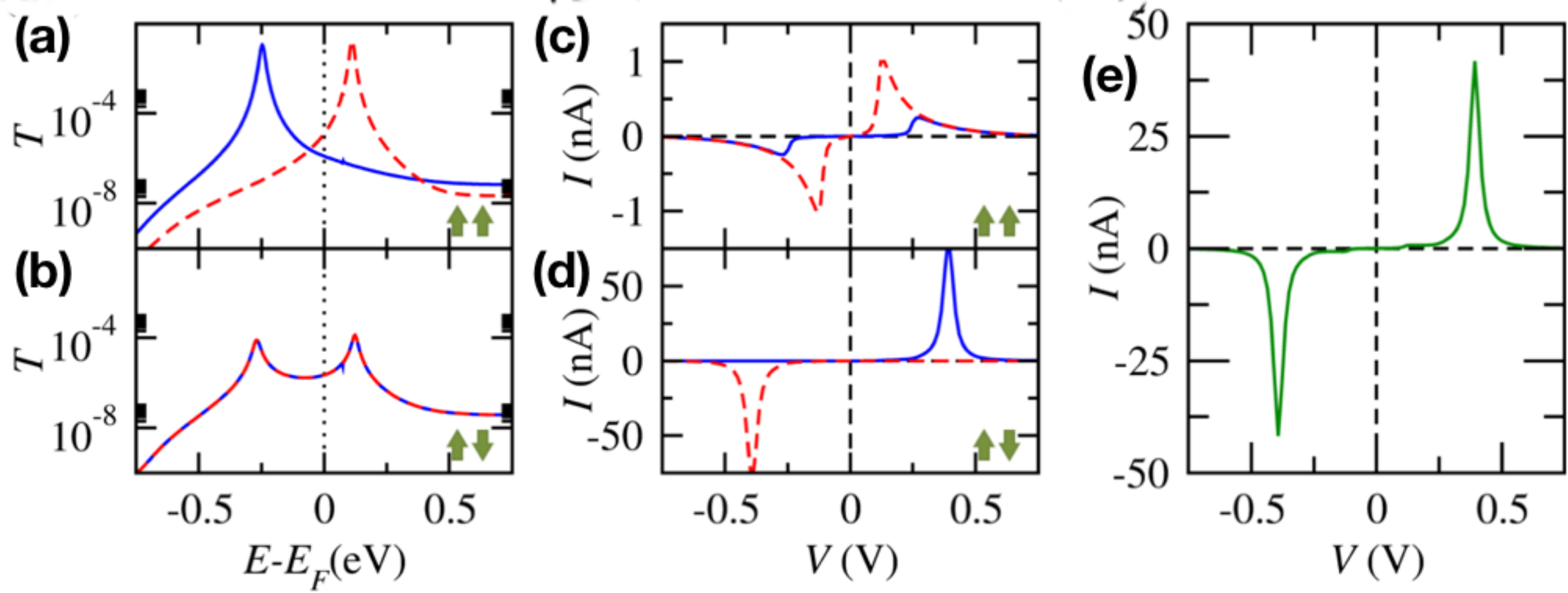}
\caption{\label{fig:WW} Electrical properties of a WW nanogap having two $R=4$ wedges facing each other. 
(a) and (b) are the spin-resolved transmission $T_\sigma(E)$ for P and AP edge spin alignments, respectively; (c) and (d) are the 
spin-polarized currents as a function of voltage for the same orientations. (e) is the total electrical current as a function of voltage.}
\end{figure*}

\subsection{Electrical properties of a WW nanogap\label{sec:WW}}
We discuss finally the electrical response of WW nanogaps where two wedges are facing each other across the gap. 
We have also simulated WW nanogaps where the two wedges are slightly shifted  relative to each other, finding results similar to those discussed 
here if the two wedges are not strictly aligned. 
We find that the zero-voltage transmission coefficients $T^{P}_\sigma(E)$ for the P wedge alignment are similar to those of a WZ nanogap, 
having resonances at energies $\epsilon_{M1,\sigma}$ (see Figure \ref{fig:WW} (a)). They correspond to tunneling between the M1 
modes at both wedges as we  show schematically in Figure \ref{fig:diagram_ws} (b) for the P spin alignment. The low-voltage current is tiny because even though the 
two levels are aligned energetically, they stay outside the integration window. As the voltage is raised from zero, the two levels de-align. 
The spin-up M1 level at the left electrode never enters the integration window, while the right spin-up M1 level does 
enter at $V=\epsilon_{M1,\uparrow}/e$, resulting in a sudden rise of the current. This rise is small though because the two spin-up levels 
are de-aligned. Further increase of the voltage results in a Negative Differential Resistance feature because even though the right spin-up M1 
level remains inside the integration window, the two levels de-align even further. The spin-resolved current-voltage  curve 
for the P alignment is shown  in Figure \ref{fig:WW} (c). 

\begin{figure*}
	\includegraphics[width=\textwidth]{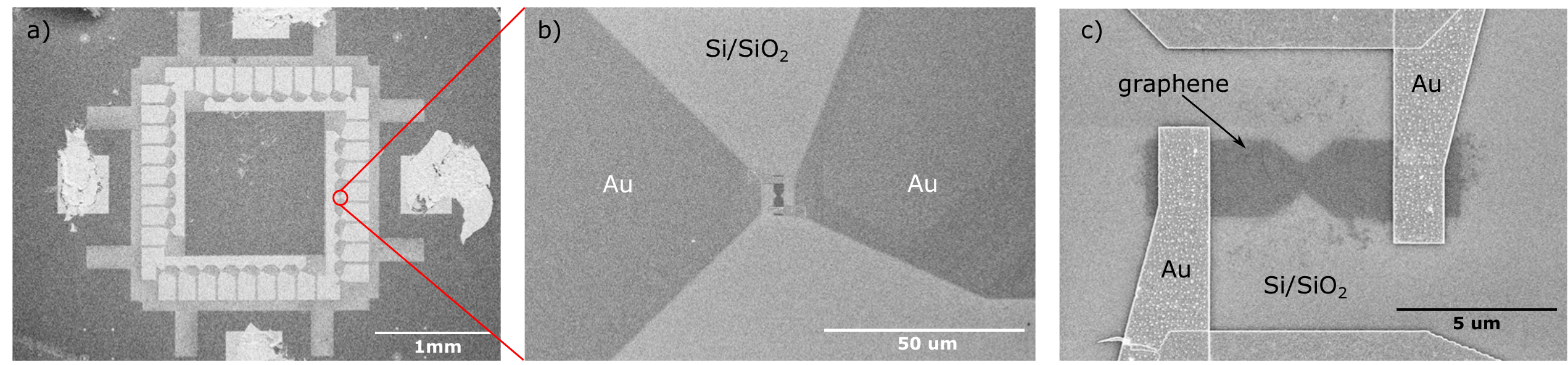}
	\caption{Scanning electron microscopy images of several graphene bridges grouped in a device. The nanogap is open in the bridge by 
	feedback controlled electroburning (see text for details).	It typically appears centered around the constriction where heat is dissipated less efficiently. }
	\label{fig:SEM}
\end{figure*}

\begin{figure*}
\includegraphics[width=\textwidth]{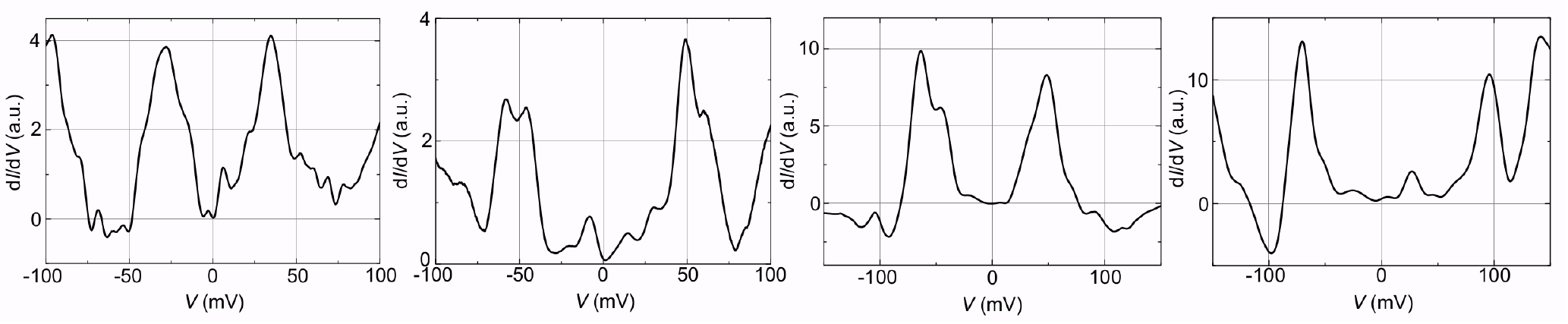}
\caption{\label{fig:experiment}
Experimental differential conductance d$I$/d$V$ measured as a function of a bias voltage  $V$ in four different graphene nanogaps fabricated by 
electroburning. Each of them show two well-defined low-voltage peaks, placed roughly symmetrically around zero voltage.}
\end{figure*}

The spin-up and -down M1 levels at the right electrode are swapped for the AP wedge alignment. As a consequence, the spin-up M1 levels 
at both wedges shift towards the Fermi level and enter the integration window when the bias voltage is increased from zero. 
Eventually they align with each other resulting in a large increase of the spin-up current. 
If the bias voltage is increased further then the 
two spin-up levels de-align and the current shrinks. The spin-down current is very small for positive biases because the spin-down 
levels at each side of the gap both move away from the integration window. Upon voltage reversal, the spin-up levels move away from 
the integration window, while the spin-down levels move towards it, eventually entering it and later on aligning with each other,
resulting in a large increase of the spin-down voltage. The computed current-voltage is shown in Figure 
\ref{fig:WW} (d). Perfect spin rectification and a large NDR signal is then expected for this nano-gap.
The current peak for the AP is one order of magnitude larger than the current plateau for the P alignment. Because the nanogap follows 
super-paramagnetic behaviour, we expect that the AP current peak should dominate the current-voltage characteristics as we show in Figure 
\ref{fig:WW} (e).

\section{Experimental measurements of graphene nanogaps}
A set of nanometer-spaced graphene electrodes is prepared is prepared by electron-beam lithography and feedback-loop controlled electroburning 
of graphene \cite{Prins2011,Burzuri2012,Island2014,Burzuri2016}. A scanning electron  microscope (SEM) image of the resulting devices is 
shown in Figure \ref{fig:SEM}.
Electron transport measurements across the gaps are performed at cryogenic temperatures ($T$ around 2K) to reduce thermal noise. In particular, 
the electrical current $I$ is measured as a function of an applied bias voltage $V$ between the two graphene electrodes. The differential 
conductance d$I$/d$V$ is thereafter numerically obtained. Some representative d$I$/d$V$ characteristics measured in four devices are 
shown in Figure \ref{fig:experiment}. These show low-voltage peaks in the 30-100 mV range, that are placed roughly symmetrically around zero voltage, however 
this symmetry is not exact. These results fit qualitatively with the theoretical differential conductance for a WZ nanogap  and 
would point to one or more WZ edges participating in the electron transport across the gap. The results are also consistent with conventional resonant
tunnelling by a single level placed inside the nanogap as shown in Figure \ref{fig:summary0} (a3).

\section{Conclusions}

We have proposed  in this article two possible realizations of graphene nano-gaps that result is spin-polarized currents originated 
by spin-polarized low-energy modes residing on the electrodes protrusions. These spin-polarized states are zero-energy modes whose
degeneracy is lifted by the Coulomb Hubbard interaction. We find that the fully spin-polarized 
might show perfect spin rectification together with low-voltage charge rectification.  Additional features, like Negative Differential 
Resistance, that is spin-polarized should be an indication that two such protrusions are roughly facing each other.
Our experimental measurements are consistent with our computational predictions for WZ nanogaps. We believe that further technological
developments will enable to achieve a detailed control of the edges shape in nanogaps, so that our proposed mechanism can be used to design
spintronically active devices.

\section{Methods}
\subsection{Technical details of the calculations \label{app:technical}}
\begin{figure}
\includegraphics[width=\columnwidth]{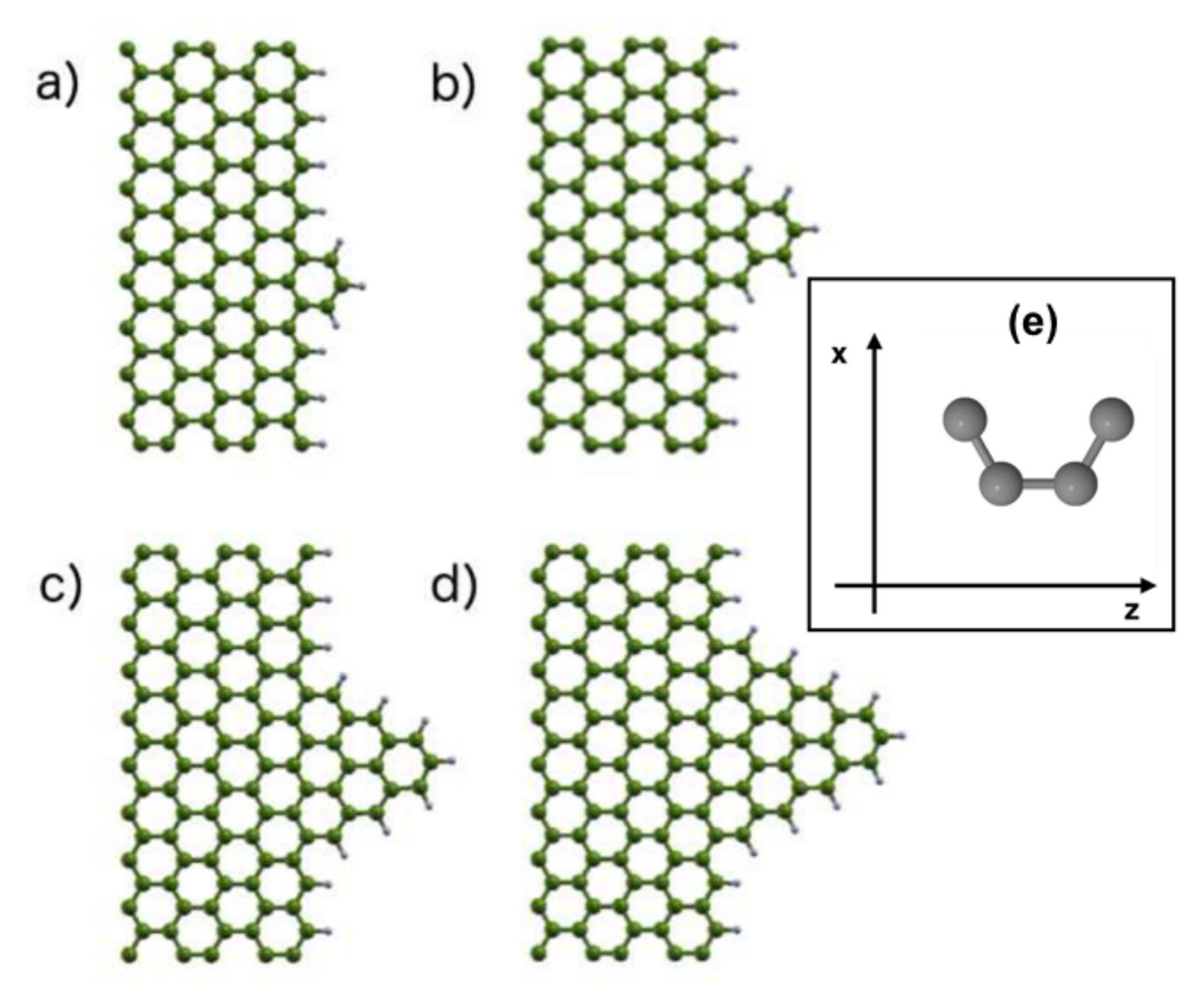}
\caption{\label{fig:wedges}
Illustration of a few possible wedges: (a) wedge containing $R=1$ row, 
having a height of $d_w=2.1\,$\AA; (b) $R=2$ rows, $d_w=4.2\,$\AA; (c) 
$R=3$ rows, $d_w=6.3\,$\AA\ and (d) $R=4$ rows, $d_w=8.4\,$\AA. The inset (e) shows the 
elementary unit cell (EUC) used to build the ribbons and electrodes of this paper. }
\end{figure}

The super-cell used  in our nanogap simulations is depicted schematically in Figure \ref{fig:wedge-example}, where  the  
X- and Z-axes are also shown. It consists of an hydrogenated ribbon whose edges run along the X-axis, are placed in the
middle of the cell, and are separated by a gap of length $d_0$. The elementary unit cell (EUC) used to build the nanogaps is
shown in Figure \ref{fig:wedges} (e) and consists of four carbon atoms. 
The height of the super-cell has $N_x=9$ EUCs that corresponds to $d_x=2.23$ nm. Periodic 
boundary conditions are applied across the three spatial directions. The super-cell can
be considered as a nanogap if $d$ is small  enough that the wave functions at the two edges overlap with each other, or a ribbon
in the opposite case.
Several possible wedges for WZ nanogaps are illustrated in 
Figure \ref{fig:wedges} (a)-(d). They are characterized by the number of extra carbon rows $R$, that correspond to a given wedge height $d_w$.
We note that the  electrical current and conductance across the gap depend exponentially on the gap length but only linearly 
on the cross-sectional length. However, because the current is dominated by the wedge, the effective tunnelling length 
$d$ determined by a Simmons' fitting \cite{S63a} of the tunnelling 
current and the apparent nanogap length $d_0$ estimated in a TEM image of the nanogap need not be the same. The strong dependence of the 
tunnelling current on $d$ conveys a large reduction of the effective  cross-sectional length $l_\text{cross}$.

We have determined the magnetic properties of SS nanogaps through the Vienna Ab 
initio Simulation Package (VASP)\cite{vasp1,vasp2}. The code expands the wave functions of the valence electrons in a 
plane-wave basis set where we have used an energy cutoff of 400 eV. Interactions among valence and core is reproduced with 
the Projector Augmented Wave (PAW) method \cite{paw1,paw2}. We have used a Generalized Gradient Approximation 
functional \cite{pbe96}. To find the optimal atomic arrangement we relaxed the inter-atomic forces using a force 
tolerance $5\times 10^{-3}$ eV/\AA. After force relaxation, we have included the Spin-Orbit interaction and have run the code again 
to determine the magnetic properties using 16 k-points along the X-axis. We have also used the 
code SIESTA to calculate magnetic moments and magnetic anisotropies\cite{Lfs06} and benchmark them against our VASP 
results. SIESTA uses norm-conserving pseudopotentials and linear combinations of atomic orbitals as basis 
set \cite{siesta}.  Because the above magnetic properties are  rather sensitive to accuracy parameters, we have used a double-zeta 
polarized basis that includes d-orbitals\cite{Gmi09}, a real-space mesh defined by a cutoff energy of 1000 Ry, an electronic 
tolerance for the density matrix of $10^{-6}$ eV and a tolerance for the maximum atomic forces of $10^{-3} \,$eV/\AA. 
We have simulated graphene islands with a triangular shape similar to the wedges. We have passivated the carbon atoms at the edges with 
hydrogen atoms or hydroxyl groups. We have used a double-zeta polarized supplemented with diffuse orbitals for all the islands' atoms,
and have relaxed all atomic forces below a tolerance of $0.02$ eV/\AA.

We have determined the electronic properties of the different nanogaps mostly via SIESTA, where we have used a double-zeta 
basis set for the inner carbon atoms at the sheets. We have described the carbon and hydrogen atoms at the wedges and edges using a 
double-zeta polarized basis set that included diffuse orbitals with quite long radii. 
We have chosen a mesh cut-off of 300 Ry to define the real space grid and the LDA exchange and correlation functional \cite{pz81}. 
We have determined the electrical response of the nanogaps via the  quantum transport code 
GOLLUM \cite{gollum}. The code reads either DFT or model tight-binding Hamiltonians to generate those transport properties, and 
we have used the DFT code SIESTA in the present article. We have checked that the above-described basis set provides a reasonable
description of the tunneling currents.  Most of the calculations discussed here do not include transverse $k-$points.

\subsection{Experimental details}
The nanogaps are fabricated on chemical vapor deposited (CVD) graphene films grown on Si/SiO2 substrates. Tens of graphene bridges narrowed 
at the central part into bow-tie constrictions ($< 1 \mu$m) are pre-patterned on the sheets by electron-beam lithography. The surrounding 
excess graphene is thereafter etched away with oxygen plasma reactive-ion etching. The electrical contact to the resulting graphene bridges 
is established through gold pads fabricated by electron-beam lithography and subsequent gold deposition. A scanning electron microscope (SEM)
 image of the resulting devices is shown in Fig. \ref{fig:SEM}. The opening of the nanogaps is achieved via a feedback-loop controlled electroburning 
technique \cite{Prins2011,Burzuri2012}. A voltage bias in the few-volts range is applied between the Au terminals at room temperature and in air. The high current 
density generated heats the flake by Joule effect until eventually some carbon atoms are removed around the constriction where the heat 
dissipation efficiency is lower \cite{Island2014,Burzuri2016}. As soon as the conductance drops around 10 \%, the voltage is ramped to zero in milliseconds to avoid 
the abrupt burning of the flake that may result in wide gaps. The burning process is repeated until the low bias ($V = 10$ mV) resistance is 
greater than 10 G$\Omega$ to avoid the presence of graphene nano-islands bridging the electrodes.

JF, AGF, VMGS and DC acknowledge funding by the Spanish MICINN through the grants FIS2012-34858 and FIS2015-63918-R.
EB thanks  funds from the EU FP7 program through Project 618082 ACMOL and the H2020-MSCA-IF 746579.
Research at Delft was supported by NOW/OCW and an ERC advanced grant (Mols@Mols; HvdZ).
Support by the  EU FP7-ITN MOLESCO grant (no. 606728) and the EU  H2020 RIA QuIET  grant (no. 767187) is also 
gratefully acknowledged.

\bibliography{master}

\end{document}